\documentstyle[twoside,fleqn,espcrc2]{article}

\title{Proton Spin Content From Lattice QCD\thanks{Talk presented by N. Mathur at Lattice '99, Pisa, Italy}}
\author{Nilmani Mathur
$^{\rm a,b}{}$, S. J. Dong$^{\rm a}{}$,
K. F. Liu\address{Department of Physics and Astronomy, 
University of Kentucky, Lexington, KY 40506-0055, USA}
and N. C. Mukhopadhyay\address{Department of Physics, Applied Physics 
and Astronomy, RPI, Troy, New York 12180-3590, USA}\\
(Kentucky Field theory Collaboration)}

\begin{document}

\begin{abstract}

We calculate the form factor of the 
quark energy momentum tensor and thereby extract 
the quark orbital angular momentum of the nucleon.
The calculation is done on a quenched $16^3 \times 24$ 
lattice at $\beta = 6.0$ and with Wilson fermions at $\kappa$ = 0.148, 
0.152, 0.154 and 0.155. We calculate the disconnected insertion 
stochastically which employs the $Z_2$ noise with an unbiased subtraction. 
This proves to be an efficient method of reduce the error from the noise.
We find that the total quark contribution to the proton spin is
$0.29 \pm 0.07$. From this we deduce that the quark orbital angular
momentum is $0.17 \pm 0.08$ and predict the gluon spin to be
$0.21 \pm 0.07$, i.e. about 40\% of the proton spin is due to the
glue. 
\end{abstract}
\maketitle
To understand the spin content of the proton remains a challenging problem 
in QCD~\cite{review}. Experimental~\cite{SMC},~\cite{SLAC} 
and lattice results~\cite{liu}, \cite{fukugita} 
suggest that the quark contribution (${1\over 2} \Sigma$) to 
the proton spin is about $25 \pm 10$\%. 
But to date, we have very little knowledge about the remaining part of the 
proton spin. We do not have reliable estimate about the spin contribution 
from the gluons or the orbital angular momentum of the quark.
In this talk, I will show our lattice results on the total 
angular momentum of the
quarks and thereby deduce the quark orbital angular momentum and predict
the gluon contribution to the proton spin.

Recently it was shown~\cite{Ji} that one can decompose the total angular 
momentum of QCD in a {\it gauge invariant} way, i.e.
\begin{eqnarray}
\vec{J}& = & \int d^{3}x \, {1 \over 2}\, \bar{\psi}\,
\vec{\gamma} \gamma_{5} \psi \,+\,\int d^{3}x \, 
\psi^{\dagger} \{\vec{x} \times (- i \vec{D})\} \psi  \nonumber\\
&&\quad\quad\quad\quad\quad\,\,+\, {\int d^{3}x \, [\,\vec{x} \times {(\vec{E} \times \vec{B})}]}.
\end{eqnarray}
The forward matrix element of this operator in the proton defines the 
decomposition of the proton spin ${1\over 2} = {1\over 2} \Sigma 
+ L_{q} + J_{g}$, 
where ${1\over 2} \Sigma$ is the quark spin contribution, $L_{q}$ is the
quark orbital angular momentum and $J_{g}$ is the total angular momentum 
of the glue. To calculate the angular momentum of the quark and the gluon
in the proton, one first notices that the gauge invariant quark-gluon 
energy momentum tensor~\cite{Ji} is
\begin{eqnarray}
T^{\mu\nu} &= & T^{\mu\nu}_{q} + T^{\mu\nu}_{g} \nonumber\\
&= & 
{1 \over 2}\,  [ \bar{\psi}\gamma^{(\mu}
{\stackrel {\longrightarrow} {iD^{\nu)}}}\psi 
+ \bar{\psi}\gamma^{(\nu}
{\stackrel {\longleftarrow} {iD^{\mu)}}}\psi ]\nonumber\\
&&\,+\,{1\over 4} g^{\mu\nu} F^{2}\,-\,F^{\mu\alpha}
F^{\nu}_{\alpha},
\end{eqnarray}
where the first part is the quark energy momentum tensor and the second 
one is that of the gluon. Form factors of this 
energy momentum tensor current can be defined as
\begin{eqnarray}
<p,s|T^{\mu\nu}_{q,g}(0)|p^{\prime},s^{\prime}>
 &=&
\bar {u}(p,s) [ {T_{1}}^{q,g}(q^{2}) \gamma^{(\mu}\bar{p}^{\nu)}
\nonumber\\
&&\hspace{-0.5in}+\,\, {T_{2}}^{q,g}(q^{2}) \bar{p}^{(\mu} i\sigma^{\nu ) \alpha}q_{\alpha}
/2m \nonumber\\
&&\hspace{-0.5in}+\,\, {1\over m} {T_{3}}^{q,g}(q^{2})
(q^{\mu}q^{\nu} -  g^{\mu\nu}
q^{2}) \nonumber\\
&&\hspace{-0.5in}\,+\, {T_{4}}^{q,g}(q^{2}) g^{\mu\nu}m ] 
u(p^{\prime},s^{\prime}),
\end{eqnarray}
where $\bar{p}^{\mu} = (p^{\mu} + {p^{\prime}}^{\mu})/2,\, 
q^{\mu} = p^{\mu} - {p^{\prime}}^{\mu}$ and $u(p)$ is the nucleon spinor.
It is proved~\cite{Ji} that the total angular momentum of the 
quark or gluon is related to the sum of the $T_1$ and $T_2$ form
factors at zero momentum transfer, i. e. 
\begin{eqnarray}
J_{q,g}&=& 
{<p,s|\,{1 \over 2}\, \epsilon^{ijk}
\int d^{3}x \,(T^{0k}_{q,g}x^{j} - T^{0j}_{q,g}x^{k})\,|p,s>
\over {<p,s|p,s>}}\nonumber\\
&=& {1 \over 2}\, [ T^{q,g}_{1}(0)\,+\,T^{q,g}_{2}(0)]
\,\,\,=\,\,\,T^{q,g}(0)\,.
\end{eqnarray}
Therefore, by calculating the form factor $T^{q,g}(q^{2})$ at different 
$q^{2}$ and then extrapolating to $q^{2} \rightarrow 0$ limit one can 
obtain the total angular momentum of the quark and the gluon separately. In 
this talk I shall present our calculation of the total angular momentum 
of the quark, $J_{q} = {1\over 2} \Sigma + L_{q}$.

To compute the matrix element, one needs to calculate the two and three 
point functions. Three point function has two parts: connected insertion (CI)
due to valence and cloud quarks, and disconnected insertion (DI) arising out 
of sea quarks. For CI, we evaluate the form factor $T(q^{2})$ in 
terms of the following ratio (for notation see ref.~\cite{ga8}):
\begin{eqnarray}
&&\hspace{-0.3in}{\hbox{Tr}\left[\Gamma_{m}G_{NT^{0 j}N}(t_{2},t_{1},\vec{0},-\vec{q})\right]
\over {\hbox{Tr}}\left[\Gamma_{e}G_{NN}(t_{2}, \vec{0})\right]}\cdot 
{{\hbox{Tr}}\left[\Gamma_{e}G_{NN}(t_{1}, \vec{0})\right] \over 
{\hbox{Tr}}\left[\Gamma_{e}G_{NN}(t_{1}, \vec{q})\right]}\nonumber\\
&&=  {1\over 2}\,\epsilon_{jkm}\, q_{k}\, T^{q,g}(q^{2})\,.
\end{eqnarray}
For CI, we calculate $T^{q}(q^{2})$ at different 
$q^{2}$ and then extrapolate that to the $q^{2} \rightarrow 0$ limit.
Results are obtained for relatively light quarks with
$\kappa\,=\,0.148, 0.152, 0.154$ and $0.155$. 
We follow the same technique used in~\cite{ga8} and~\cite{liu}. 
Fig.1 shows the dipole fitting of $T(q^{2})$ at different $q^{2}$ for 
$\kappa = 0.152$ and 0.155. Plots for the other $\kappa$'s are similar. 
\vspace*{-0.52in}
\begin{figure}[h]
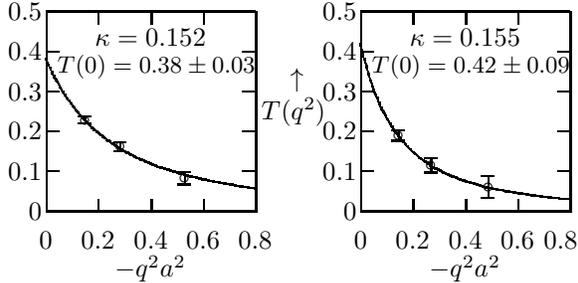

\[
\hspace*{-0.4in}
\input{lat_pole_c52.tex}
\hspace*{-0.25in}
\input{lat_pole_c55.tex}
\]
\vspace*{-0.52in}
\caption{Dipole fitting for CI. Extrapolated ($q^{2}\rightarrow 0$) values 
are shown by $T_{L}(0)$.}
\end{figure}
\vspace*{-0.2in}

The chiral limit for  $J_{q}^{CI}$ is 
taken with a linear dependence on the quark mass $m_{q}a$ for 
these four $\kappa$ (fig.3). To account for 
the correlations, both fittings in $q^2$ and $m_q$ are 
done by using the covariant matrix and the final error at the chiral limit 
is obtained from the Jackknife procedure. Finally, to get the continuum 
result we multiply the lattice $T_{L}(0)$ by the tadpole improved 
renormalization constant~\cite{cap} for the operator $T^{0j}$ and obtain 
the CI part of the total angular momentum of quarks $J^{CI}_{q} = 
0.43 \pm 0.07$, which almost saturates the proton spin.
The calculation is done on a quenched $16^3 \times 24$ lattice 
at $\beta = 6.0$ with Wilson fermions and with 100 configurations. 
For DI contribution, we use 
the relation~\cite{discon},
\begin{eqnarray}  \label{DI}
{1\over 3} \sum^{3}_{i = 1}\sum_{\tau}{\Gamma^{\beta\alpha}_{i} 
G^{\alpha\beta}_{PT_{4i}P}(t_{f} \tau) \over G^{\alpha\beta}_{PP}(t_{f})} 
&&\hspace{-0.2in}{\stackrel{t_{f}>>a}\longrightarrow} C
+ t_{f}\,.\,T^{L}_{DI}\quad\,\,
\end{eqnarray}
where $C$ is a constant and DI is obtained from the slope of the sum on the left hand side as a 
function of $t_{f}$. To calculate the trace of the three point function in 
DI, we follow the same methodology as in ref.~\cite{liu} and use the same stochastic algorithm, 
which employs 
the $Z_2$ noise estimator and provides the estimate with minimum variance. 
In addition, we use two more techniques to reduce the errors. First one 
is the removal of the unwanted real or imaginary part of the three point 
function through charge conjugation and hermiticity (CH) symmetry. 
Secondly, we use an unbiased subtraction method~\cite{chris} 
to reduce the variance due to the stochastic estimator, i. e. we use the
following estimator 
\begin{equation}
\hbox{Tr}\left( A^{-1} \right)
= E {[< \eta^{\dagger} ( A^{-1} 
- \sum^{P}_{i=1} \lambda_{i}\, O^{(i)} ) \eta > ]},
\end{equation}
where $\eta$'s are $Z_{2}$ noise vectors, $O^{(i)}$'s are a set of $P$ 
traceless matrices and $\lambda_{i}$'s are variational constants which 
need to be tuned to reduce the variance of Tr$(A^{-1})$.
In principle, one can take any set of traceless matrices as a choice of 
$O^{(i)}$. But we choose a set of traceless matrices obtained from the 
hoping parameter expansion of the propagator in order to match the 
off-diagonal behavior of the matrix $A^{-1}$ so that they can cancel the 
off-diagonal contributions to the variance~\cite{chris}. A variational 
program which minimizes the variance from the 100 gauge configurations is 
used to determine the optimum set of $\lambda$'s. 
With these two techniques,
we obtain a reduction of 
error by as much as 3-4 times. The subtraction method is also proved to 
be very efficient for the calculations of fermion determinants~\cite{chris} 
and for the traces of many other operators~\cite{walter}. 
Fig.2 shows a plot of sum {\it vs} $t_{f}$ (Eq.(\ref{DI})) for 
$\kappa = 0.154$ and $q^{2} = 1$ with and without subtraction. 
Finally, we fit $T^{L}_{dis}$ as a function of $q^{2}$ 
by a monopole form which is shown in Fig.2 for $\kappa = 0.154$. 
Similar to CI, we also use covariant matrix fitting and the final error-bar 
is obtained by the jackknife method. A finite mass correction 
factor~\cite{lagae} is introduced while extrapolating to the chiral limit.
For strange quark contribution we follow the same procedure as in ref.~\cite{liu}.
\vspace*{-0.55in}
\begin{figure}[h]
\[
\hspace*{-0.36in}
\input{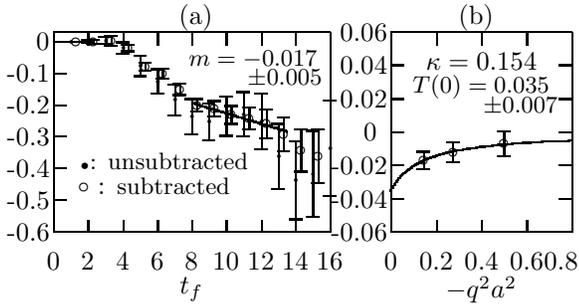}
\hspace*{-0.48in}
\setlength{\unitlength}{0.240900pt}
\ifx\plotpoint\undefined\newsavebox{\plotpoint}\fi
\sbox{\plotpoint}{\rule[-0.200pt]{0.400pt}{0.400pt}}%
\begin{picture}(524,495)(0,0)
\font\gnuplot=cmr10 at 10pt
\gnuplot
\sbox{\plotpoint}{\rule[-0.200pt]{0.400pt}{0.400pt}}%
\put(176.0,113.0){\rule[-0.200pt]{4.818pt}{0.400pt}}
\put(170,113){\makebox(0,0)[r]{-0.06}}
\put(440.0,113.0){\rule[-0.200pt]{4.818pt}{0.400pt}}
\put(176.0,165.0){\rule[-0.200pt]{4.818pt}{0.400pt}}
\put(170,165){\makebox(0,0)[r]{-0.04}}
\put(440.0,165.0){\rule[-0.200pt]{4.818pt}{0.400pt}}
\put(176.0,218.0){\rule[-0.200pt]{4.818pt}{0.400pt}}
\put(170,218){\makebox(0,0)[r]{-0.02}}
\put(440.0,218.0){\rule[-0.200pt]{4.818pt}{0.400pt}}
\put(176.0,270.0){\rule[-0.200pt]{4.818pt}{0.400pt}}
\put(154,270){\makebox(0,0)[r]{0}}
\put(440.0,270.0){\rule[-0.200pt]{4.818pt}{0.400pt}}
\put(176.0,322.0){\rule[-0.200pt]{4.818pt}{0.400pt}}
\put(170,322){\makebox(0,0)[r]{0.02}}
\put(440.0,322.0){\rule[-0.200pt]{4.818pt}{0.400pt}}
\put(176.0,375.0){\rule[-0.200pt]{4.818pt}{0.400pt}}
\put(170,375){\makebox(0,0)[r]{0.04}}
\put(440.0,375.0){\rule[-0.200pt]{4.818pt}{0.400pt}}
\put(176.0,427.0){\rule[-0.200pt]{4.818pt}{0.400pt}}
\put(170,427){\makebox(0,0)[r]{0.06}}
\put(440.0,427.0){\rule[-0.200pt]{4.818pt}{0.400pt}}
\put(176.0,113.0){\rule[-0.200pt]{0.400pt}{4.818pt}}
\put(176,68){\makebox(0,0){0}}
\put(176.0,407.0){\rule[-0.200pt]{0.400pt}{4.818pt}}
\put(247.0,113.0){\rule[-0.200pt]{0.400pt}{4.818pt}}
\put(247,68){\makebox(0,0){0.2}}
\put(247.0,407.0){\rule[-0.200pt]{0.400pt}{4.818pt}}
\put(318.0,113.0){\rule[-0.200pt]{0.400pt}{4.818pt}}
\put(318,68){\makebox(0,0){0.4}}
\put(318.0,407.0){\rule[-0.200pt]{0.400pt}{4.818pt}}
\put(389.0,113.0){\rule[-0.200pt]{0.400pt}{4.818pt}}
\put(389,68){\makebox(0,0){0.6}}
\put(389.0,407.0){\rule[-0.200pt]{0.400pt}{4.818pt}}
\put(460.0,113.0){\rule[-0.200pt]{0.400pt}{4.818pt}}
\put(443,68){\makebox(0,0){0.8}}
\put(460.0,407.0){\rule[-0.200pt]{0.400pt}{4.818pt}}
\put(176.0,113.0){\rule[-0.200pt]{68.416pt}{0.400pt}}
\put(460.0,113.0){\rule[-0.200pt]{0.400pt}{75.643pt}}
\put(176.0,427.0){\rule[-0.200pt]{68.416pt}{0.400pt}}
\put(318,23){\makebox(0,0){{$-q^{2}a^{2}$}}}
\put(310,450){\makebox(0,0){{(b)}}}    
\put(318,382){\makebox(0,0){{$\kappa = 0.154$}}}
\put(318,342){\makebox(0,0){{\small$T(0)=0.035$}}}
\put(380,312){\makebox(0,0){{\small$\pm0.007$}}}
\put(176.0,113.0){\rule[-0.200pt]{0.400pt}{75.643pt}}
\put(227,226){\circle{12}}
\put(273,239){\circle{12}}
\put(353,252){\circle{12}}
\put(227.0,212.0){\rule[-0.200pt]{0.400pt}{6.504pt}}
\put(217.0,212.0){\rule[-0.200pt]{4.818pt}{0.400pt}}
\put(217.0,239.0){\rule[-0.200pt]{4.818pt}{0.400pt}}
\put(273.0,223.0){\rule[-0.200pt]{0.400pt}{7.468pt}}
\put(263.0,223.0){\rule[-0.200pt]{4.818pt}{0.400pt}}
\put(263.0,254.0){\rule[-0.200pt]{4.818pt}{0.400pt}}
\put(353.0,232.0){\rule[-0.200pt]{0.400pt}{9.395pt}}
\put(343.0,232.0){\rule[-0.200pt]{4.818pt}{0.400pt}}
\put(343.0,271.0){\rule[-0.200pt]{4.818pt}{0.400pt}}
\put(176,177){\usebox{\plotpoint}}
\multiput(176.61,177.00)(0.447,1.132){3}{\rule{0.108pt}{0.900pt}}
\multiput(175.17,177.00)(3.000,4.132){2}{\rule{0.400pt}{0.450pt}}
\multiput(179.61,183.00)(0.447,0.685){3}{\rule{0.108pt}{0.633pt}}
\multiput(178.17,183.00)(3.000,2.685){2}{\rule{0.400pt}{0.317pt}}
\multiput(182.61,187.00)(0.447,0.909){3}{\rule{0.108pt}{0.767pt}}
\multiput(181.17,187.00)(3.000,3.409){2}{\rule{0.400pt}{0.383pt}}
\put(185.17,192){\rule{0.400pt}{0.700pt}}
\multiput(184.17,192.00)(2.000,1.547){2}{\rule{0.400pt}{0.350pt}}
\multiput(187.61,195.00)(0.447,0.685){3}{\rule{0.108pt}{0.633pt}}
\multiput(186.17,195.00)(3.000,2.685){2}{\rule{0.400pt}{0.317pt}}
\multiput(190.00,199.61)(0.462,0.447){3}{\rule{0.500pt}{0.108pt}}
\multiput(190.00,198.17)(1.962,3.000){2}{\rule{0.250pt}{0.400pt}}
\multiput(193.00,202.61)(0.462,0.447){3}{\rule{0.500pt}{0.108pt}}
\multiput(193.00,201.17)(1.962,3.000){2}{\rule{0.250pt}{0.400pt}}
\put(196,205.17){\rule{0.700pt}{0.400pt}}
\multiput(196.00,204.17)(1.547,2.000){2}{\rule{0.350pt}{0.400pt}}
\multiput(199.00,207.61)(0.462,0.447){3}{\rule{0.500pt}{0.108pt}}
\multiput(199.00,206.17)(1.962,3.000){2}{\rule{0.250pt}{0.400pt}}
\put(202,210.17){\rule{0.700pt}{0.400pt}}
\multiput(202.00,209.17)(1.547,2.000){2}{\rule{0.350pt}{0.400pt}}
\put(205,212.17){\rule{0.700pt}{0.400pt}}
\multiput(205.00,211.17)(1.547,2.000){2}{\rule{0.350pt}{0.400pt}}
\put(208,214.17){\rule{0.482pt}{0.400pt}}
\multiput(208.00,213.17)(1.000,2.000){2}{\rule{0.241pt}{0.400pt}}
\put(210,216.17){\rule{0.700pt}{0.400pt}}
\multiput(210.00,215.17)(1.547,2.000){2}{\rule{0.350pt}{0.400pt}}
\put(213,218.17){\rule{0.700pt}{0.400pt}}
\multiput(213.00,217.17)(1.547,2.000){2}{\rule{0.350pt}{0.400pt}}
\put(216,219.67){\rule{0.723pt}{0.400pt}}
\multiput(216.00,219.17)(1.500,1.000){2}{\rule{0.361pt}{0.400pt}}
\put(219,221.17){\rule{0.700pt}{0.400pt}}
\multiput(219.00,220.17)(1.547,2.000){2}{\rule{0.350pt}{0.400pt}}
\put(222,222.67){\rule{0.723pt}{0.400pt}}
\multiput(222.00,222.17)(1.500,1.000){2}{\rule{0.361pt}{0.400pt}}
\put(225,223.67){\rule{0.723pt}{0.400pt}}
\multiput(225.00,223.17)(1.500,1.000){2}{\rule{0.361pt}{0.400pt}}
\put(228,225.17){\rule{0.700pt}{0.400pt}}
\multiput(228.00,224.17)(1.547,2.000){2}{\rule{0.350pt}{0.400pt}}
\put(231,226.67){\rule{0.482pt}{0.400pt}}
\multiput(231.00,226.17)(1.000,1.000){2}{\rule{0.241pt}{0.400pt}}
\put(233,227.67){\rule{0.723pt}{0.400pt}}
\multiput(233.00,227.17)(1.500,1.000){2}{\rule{0.361pt}{0.400pt}}
\put(236,228.67){\rule{0.723pt}{0.400pt}}
\multiput(236.00,228.17)(1.500,1.000){2}{\rule{0.361pt}{0.400pt}}
\put(239,229.67){\rule{0.723pt}{0.400pt}}
\multiput(239.00,229.17)(1.500,1.000){2}{\rule{0.361pt}{0.400pt}}
\put(242,230.67){\rule{0.723pt}{0.400pt}}
\multiput(242.00,230.17)(1.500,1.000){2}{\rule{0.361pt}{0.400pt}}
\put(245,231.67){\rule{0.723pt}{0.400pt}}
\multiput(245.00,231.17)(1.500,1.000){2}{\rule{0.361pt}{0.400pt}}
\put(248,232.67){\rule{0.723pt}{0.400pt}}
\multiput(248.00,232.17)(1.500,1.000){2}{\rule{0.361pt}{0.400pt}}
\put(251,233.67){\rule{0.482pt}{0.400pt}}
\multiput(251.00,233.17)(1.000,1.000){2}{\rule{0.241pt}{0.400pt}}
\put(256,234.67){\rule{0.723pt}{0.400pt}}
\multiput(256.00,234.17)(1.500,1.000){2}{\rule{0.361pt}{0.400pt}}
\put(259,235.67){\rule{0.723pt}{0.400pt}}
\multiput(259.00,235.17)(1.500,1.000){2}{\rule{0.361pt}{0.400pt}}
\put(262,236.67){\rule{0.723pt}{0.400pt}}
\multiput(262.00,236.17)(1.500,1.000){2}{\rule{0.361pt}{0.400pt}}
\put(253.0,235.0){\rule[-0.200pt]{0.723pt}{0.400pt}}
\put(268,237.67){\rule{0.723pt}{0.400pt}}
\multiput(268.00,237.17)(1.500,1.000){2}{\rule{0.361pt}{0.400pt}}
\put(271,238.67){\rule{0.723pt}{0.400pt}}
\multiput(271.00,238.17)(1.500,1.000){2}{\rule{0.361pt}{0.400pt}}
\put(265.0,238.0){\rule[-0.200pt]{0.723pt}{0.400pt}}
\put(276,239.67){\rule{0.723pt}{0.400pt}}
\multiput(276.00,239.17)(1.500,1.000){2}{\rule{0.361pt}{0.400pt}}
\put(274.0,240.0){\rule[-0.200pt]{0.482pt}{0.400pt}}
\put(282,240.67){\rule{0.723pt}{0.400pt}}
\multiput(282.00,240.17)(1.500,1.000){2}{\rule{0.361pt}{0.400pt}}
\put(279.0,241.0){\rule[-0.200pt]{0.723pt}{0.400pt}}
\put(288,241.67){\rule{0.723pt}{0.400pt}}
\multiput(288.00,241.17)(1.500,1.000){2}{\rule{0.361pt}{0.400pt}}
\put(285.0,242.0){\rule[-0.200pt]{0.723pt}{0.400pt}}
\put(294,242.67){\rule{0.482pt}{0.400pt}}
\multiput(294.00,242.17)(1.000,1.000){2}{\rule{0.241pt}{0.400pt}}
\put(291.0,243.0){\rule[-0.200pt]{0.723pt}{0.400pt}}
\put(299,243.67){\rule{0.723pt}{0.400pt}}
\multiput(299.00,243.17)(1.500,1.000){2}{\rule{0.361pt}{0.400pt}}
\put(296.0,244.0){\rule[-0.200pt]{0.723pt}{0.400pt}}
\put(308,244.67){\rule{0.723pt}{0.400pt}}
\multiput(308.00,244.17)(1.500,1.000){2}{\rule{0.361pt}{0.400pt}}
\put(302.0,245.0){\rule[-0.200pt]{1.445pt}{0.400pt}}
\put(314,245.67){\rule{0.723pt}{0.400pt}}
\multiput(314.00,245.17)(1.500,1.000){2}{\rule{0.361pt}{0.400pt}}
\put(311.0,246.0){\rule[-0.200pt]{0.723pt}{0.400pt}}
\put(322,246.67){\rule{0.723pt}{0.400pt}}
\multiput(322.00,246.17)(1.500,1.000){2}{\rule{0.361pt}{0.400pt}}
\put(317.0,247.0){\rule[-0.200pt]{1.204pt}{0.400pt}}
\put(334,247.67){\rule{0.723pt}{0.400pt}}
\multiput(334.00,247.17)(1.500,1.000){2}{\rule{0.361pt}{0.400pt}}
\put(325.0,248.0){\rule[-0.200pt]{2.168pt}{0.400pt}}
\put(342,248.67){\rule{0.723pt}{0.400pt}}
\multiput(342.00,248.17)(1.500,1.000){2}{\rule{0.361pt}{0.400pt}}
\put(337.0,249.0){\rule[-0.200pt]{1.204pt}{0.400pt}}
\put(354,249.67){\rule{0.723pt}{0.400pt}}
\multiput(354.00,249.17)(1.500,1.000){2}{\rule{0.361pt}{0.400pt}}
\put(345.0,250.0){\rule[-0.200pt]{2.168pt}{0.400pt}}
\put(365,250.67){\rule{0.723pt}{0.400pt}}
\multiput(365.00,250.17)(1.500,1.000){2}{\rule{0.361pt}{0.400pt}}
\put(357.0,251.0){\rule[-0.200pt]{1.927pt}{0.400pt}}
\put(380,251.67){\rule{0.723pt}{0.400pt}}
\multiput(380.00,251.17)(1.500,1.000){2}{\rule{0.361pt}{0.400pt}}
\put(368.0,252.0){\rule[-0.200pt]{2.891pt}{0.400pt}}
\put(394,252.67){\rule{0.723pt}{0.400pt}}
\multiput(394.00,252.17)(1.500,1.000){2}{\rule{0.361pt}{0.400pt}}
\put(383.0,253.0){\rule[-0.200pt]{2.650pt}{0.400pt}}
\put(411,253.67){\rule{0.723pt}{0.400pt}}
\multiput(411.00,253.17)(1.500,1.000){2}{\rule{0.361pt}{0.400pt}}
\put(397.0,254.0){\rule[-0.200pt]{3.373pt}{0.400pt}}
\put(431,254.67){\rule{0.723pt}{0.400pt}}
\multiput(431.00,254.17)(1.500,1.000){2}{\rule{0.361pt}{0.400pt}}
\put(414.0,255.0){\rule[-0.200pt]{4.095pt}{0.400pt}}
\put(454,255.67){\rule{0.723pt}{0.400pt}}
\multiput(454.00,255.17)(1.500,1.000){2}{\rule{0.361pt}{0.400pt}}
\put(434.0,256.0){\rule[-0.200pt]{4.818pt}{0.400pt}}
\put(457.0,257.0){\rule[-0.200pt]{0.723pt}{0.400pt}}
\end{picture}
\]
\vspace*{-0.55in}
\caption{(a) The ratio of Eq.(\ref{DI}) for $\kappa = 0.154$ at $q^{2}=1$.
 Ratios with subtraction are shifted a little towards right. $m$ is the slope.
(b) The monopole fitting of $T(q^2)_{dis}$ for several 
$q^{2}$.}
\end{figure}
\vspace*{-0.85in}
\begin{figure}[h]
\[
\hspace*{-0.37in}
\setlength{\unitlength}{0.240900pt}
\ifx\plotpoint\undefined\newsavebox{\plotpoint}\fi
\sbox{\plotpoint}{\rule[-0.200pt]{0.400pt}{0.400pt}}%
\begin{picture}(569,450)(0,0)
\font\gnuplot=cmr10 at 10pt
\gnuplot
\sbox{\plotpoint}{\rule[-0.200pt]{0.400pt}{0.400pt}}%
\put(176.0,113.0){\rule[-0.200pt]{4.818pt}{0.400pt}}
\put(170,113){\makebox(0,0)[r]{0}}
\put(485.0,113.0){\rule[-0.200pt]{4.818pt}{0.400pt}}
\put(176.0,192.0){\rule[-0.200pt]{4.818pt}{0.400pt}}
\put(170,192){\makebox(0,0)[r]{0.2}}
\put(485.0,192.0){\rule[-0.200pt]{4.818pt}{0.400pt}}
\put(176.0,270.0){\rule[-0.200pt]{4.818pt}{0.400pt}}
\put(170,270){\makebox(0,0)[r]{0.4}}
\put(485.0,270.0){\rule[-0.200pt]{4.818pt}{0.400pt}}
\put(176.0,349.0){\rule[-0.200pt]{4.818pt}{0.400pt}}
\put(170,349){\makebox(0,0)[r]{0.6}}
\put(485.0,349.0){\rule[-0.200pt]{4.818pt}{0.400pt}}
\put(176.0,427.0){\rule[-0.200pt]{4.818pt}{0.400pt}}
\put(170,427){\makebox(0,0)[r]{0.8}}
\put(485.0,427.0){\rule[-0.200pt]{4.818pt}{0.400pt}}
\put(176.0,113.0){\rule[-0.200pt]{0.400pt}{4.818pt}}
\put(176,68){\makebox(0,0){0}}
\put(176.0,407.0){\rule[-0.200pt]{0.400pt}{4.818pt}}
\put(286.0,113.0){\rule[-0.200pt]{0.400pt}{4.818pt}}
\put(286,68){\makebox(0,0){0.1}}
\put(286.0,407.0){\rule[-0.200pt]{0.400pt}{4.818pt}}
\put(395.0,113.0){\rule[-0.200pt]{0.400pt}{4.818pt}}
\put(395,68){\makebox(0,0){0.2}}
\put(395.0,407.0){\rule[-0.200pt]{0.400pt}{4.818pt}}
\put(505.0,113.0){\rule[-0.200pt]{0.400pt}{4.818pt}}
\put(505,68){\makebox(0,0){0.3}}
\put(505.0,407.0){\rule[-0.200pt]{0.400pt}{4.818pt}}
\put(176.0,113.0){\rule[-0.200pt]{79.256pt}{0.400pt}}
\put(505.0,113.0){\rule[-0.200pt]{0.400pt}{75.643pt}}
\put(176.0,427.0){\rule[-0.200pt]{79.256pt}{0.400pt}}
\put(340,23){\makebox(0,0){{$m_{q}a = ln(4\kappa_{c}/\kappa-3)$}}}
\put(176.0,113.0){\rule[-0.200pt]{0.400pt}{75.643pt}}
\put(176,275){\circle{12}}
\put(167,265){$\bullet$}
\put(224,269){\circle{12}}
\put(251,268){\circle{12}}
\put(304,266){\circle{12}}
\put(408,257){\circle{12}}
\put(337,450){\makebox(0,0){{{{CI}}}}} 
\put(176.0,247.0){\rule[-0.200pt]{0.400pt}{13.249pt}}
\put(166.0,247.0){\rule[-0.200pt]{4.818pt}{0.400pt}}
\put(166.0,302.0){\rule[-0.200pt]{4.818pt}{0.400pt}}
\put(224.0,240.0){\rule[-0.200pt]{0.400pt}{14.213pt}}
\put(214.0,240.0){\rule[-0.200pt]{4.818pt}{0.400pt}}
\put(214.0,299.0){\rule[-0.200pt]{4.818pt}{0.400pt}}
\put(251.0,249.0){\rule[-0.200pt]{0.400pt}{9.154pt}}
\put(241.0,249.0){\rule[-0.200pt]{4.818pt}{0.400pt}}
\put(241.0,287.0){\rule[-0.200pt]{4.818pt}{0.400pt}}
\put(304.0,254.0){\rule[-0.200pt]{0.400pt}{5.782pt}}
\put(294.0,254.0){\rule[-0.200pt]{4.818pt}{0.400pt}}
\put(294.0,278.0){\rule[-0.200pt]{4.818pt}{0.400pt}}
\put(408.0,248.0){\rule[-0.200pt]{0.400pt}{4.336pt}}
\put(398.0,248.0){\rule[-0.200pt]{4.818pt}{0.400pt}}
\put(398.0,266.0){\rule[-0.200pt]{4.818pt}{0.400pt}}
\put(176,274){\usebox{\plotpoint}}
\put(186,272.67){\rule{0.723pt}{0.400pt}}
\multiput(186.00,273.17)(1.500,-1.000){2}{\rule{0.361pt}{0.400pt}}
\put(176.0,274.0){\rule[-0.200pt]{2.409pt}{0.400pt}}
\put(199,271.67){\rule{0.964pt}{0.400pt}}
\multiput(199.00,272.17)(2.000,-1.000){2}{\rule{0.482pt}{0.400pt}}
\put(189.0,273.0){\rule[-0.200pt]{2.409pt}{0.400pt}}
\put(216,270.67){\rule{0.723pt}{0.400pt}}
\multiput(216.00,271.17)(1.500,-1.000){2}{\rule{0.361pt}{0.400pt}}
\put(203.0,272.0){\rule[-0.200pt]{3.132pt}{0.400pt}}
\put(229,269.67){\rule{0.723pt}{0.400pt}}
\multiput(229.00,270.17)(1.500,-1.000){2}{\rule{0.361pt}{0.400pt}}
\put(219.0,271.0){\rule[-0.200pt]{2.409pt}{0.400pt}}
\put(242,268.67){\rule{0.964pt}{0.400pt}}
\multiput(242.00,269.17)(2.000,-1.000){2}{\rule{0.482pt}{0.400pt}}
\put(232.0,270.0){\rule[-0.200pt]{2.409pt}{0.400pt}}
\put(256,267.67){\rule{0.723pt}{0.400pt}}
\multiput(256.00,268.17)(1.500,-1.000){2}{\rule{0.361pt}{0.400pt}}
\put(246.0,269.0){\rule[-0.200pt]{2.409pt}{0.400pt}}
\put(269,266.67){\rule{0.723pt}{0.400pt}}
\multiput(269.00,267.17)(1.500,-1.000){2}{\rule{0.361pt}{0.400pt}}
\put(259.0,268.0){\rule[-0.200pt]{2.409pt}{0.400pt}}
\put(282,265.67){\rule{0.964pt}{0.400pt}}
\multiput(282.00,266.17)(2.000,-1.000){2}{\rule{0.482pt}{0.400pt}}
\put(272.0,267.0){\rule[-0.200pt]{2.409pt}{0.400pt}}
\put(296,264.67){\rule{0.723pt}{0.400pt}}
\multiput(296.00,265.17)(1.500,-1.000){2}{\rule{0.361pt}{0.400pt}}
\put(286.0,266.0){\rule[-0.200pt]{2.409pt}{0.400pt}}
\put(309,263.67){\rule{0.723pt}{0.400pt}}
\multiput(309.00,264.17)(1.500,-1.000){2}{\rule{0.361pt}{0.400pt}}
\put(299.0,265.0){\rule[-0.200pt]{2.409pt}{0.400pt}}
\put(322,262.67){\rule{0.964pt}{0.400pt}}
\multiput(322.00,263.17)(2.000,-1.000){2}{\rule{0.482pt}{0.400pt}}
\put(312.0,264.0){\rule[-0.200pt]{2.409pt}{0.400pt}}
\put(336,261.67){\rule{0.723pt}{0.400pt}}
\multiput(336.00,262.17)(1.500,-1.000){2}{\rule{0.361pt}{0.400pt}}
\put(326.0,263.0){\rule[-0.200pt]{2.409pt}{0.400pt}}
\put(349,260.67){\rule{0.723pt}{0.400pt}}
\multiput(349.00,261.17)(1.500,-1.000){2}{\rule{0.361pt}{0.400pt}}
\put(339.0,262.0){\rule[-0.200pt]{2.409pt}{0.400pt}}
\put(362,259.67){\rule{0.723pt}{0.400pt}}
\multiput(362.00,260.17)(1.500,-1.000){2}{\rule{0.361pt}{0.400pt}}
\put(352.0,261.0){\rule[-0.200pt]{2.409pt}{0.400pt}}
\put(379,258.67){\rule{0.723pt}{0.400pt}}
\multiput(379.00,259.17)(1.500,-1.000){2}{\rule{0.361pt}{0.400pt}}
\put(365.0,260.0){\rule[-0.200pt]{3.373pt}{0.400pt}}
\put(392,257.67){\rule{0.723pt}{0.400pt}}
\multiput(392.00,258.17)(1.500,-1.000){2}{\rule{0.361pt}{0.400pt}}
\put(382.0,259.0){\rule[-0.200pt]{2.409pt}{0.400pt}}
\put(405,256.67){\rule{0.964pt}{0.400pt}}
\multiput(405.00,257.17)(2.000,-1.000){2}{\rule{0.482pt}{0.400pt}}
\put(395.0,258.0){\rule[-0.200pt]{2.409pt}{0.400pt}}
\put(419,255.67){\rule{0.723pt}{0.400pt}}
\multiput(419.00,256.17)(1.500,-1.000){2}{\rule{0.361pt}{0.400pt}}
\put(409.0,257.0){\rule[-0.200pt]{2.409pt}{0.400pt}}
\put(432,254.67){\rule{0.723pt}{0.400pt}}
\multiput(432.00,255.17)(1.500,-1.000){2}{\rule{0.361pt}{0.400pt}}
\put(422.0,256.0){\rule[-0.200pt]{2.409pt}{0.400pt}}
\put(445,253.67){\rule{0.964pt}{0.400pt}}
\multiput(445.00,254.17)(2.000,-1.000){2}{\rule{0.482pt}{0.400pt}}
\put(435.0,255.0){\rule[-0.200pt]{2.409pt}{0.400pt}}
\put(458,252.67){\rule{0.964pt}{0.400pt}}
\multiput(458.00,253.17)(2.000,-1.000){2}{\rule{0.482pt}{0.400pt}}
\put(449.0,254.0){\rule[-0.200pt]{2.168pt}{0.400pt}}
\put(472,251.67){\rule{0.723pt}{0.400pt}}
\multiput(472.00,252.17)(1.500,-1.000){2}{\rule{0.361pt}{0.400pt}}
\put(462.0,253.0){\rule[-0.200pt]{2.409pt}{0.400pt}}
\put(485,250.67){\rule{0.723pt}{0.400pt}}
\multiput(485.00,251.17)(1.500,-1.000){2}{\rule{0.361pt}{0.400pt}}
\put(475.0,252.0){\rule[-0.200pt]{2.409pt}{0.400pt}}
\put(498,249.67){\rule{0.964pt}{0.400pt}}
\multiput(498.00,250.17)(2.000,-1.000){2}{\rule{0.482pt}{0.400pt}}
\put(488.0,251.0){\rule[-0.200pt]{2.409pt}{0.400pt}}
\put(502.0,250.0){\rule[-0.200pt]{0.723pt}{0.400pt}}
\end{picture}
\hspace*{-0.2in}
\input{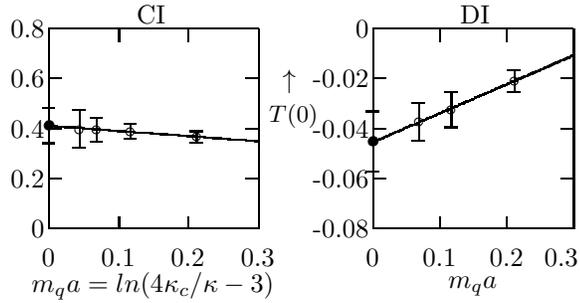}
\]
\vspace*{-0.55in}
\caption{The lattice $T_{CI}^{L}(0)$ and $T_{DI}^{L}(0)$ for CI and DI as 
a function of quark mass $ma$. The chiral result is indicated by $\bullet$.}
\end{figure}

\vspace*{-0.3in}
From DI calculation, we find that 
$J^{DI}_{q}$, like the DI part of the quark spin
${1\over 2} \Sigma^{DI}$, is also flavor symmetric, {\it i.e.,} 
$J^{DI}_{u} = J^{DI}_{d} \simeq J^{DI}_{s} 
= -0.047 \pm 0.013$. Therefore, the total DI contribution 
$J^{DI}_{q} = -0.14 \pm 0.04$. Here one should note that in 
ref.~\cite{liu}, results for ${1\over 2} \Sigma ^{DI} = -0.18 \pm 0.03$, 
which suggests that {\it the sea quarks give very little orbital angular 
momentum contribution}. 
Adding CI and DI contributions, we obtain $J_{q} = 0.29 \pm 0.07$ and 
therefore we can predict the gluon angular momentum content from the
spin sum rule, {\it i.e.,} $J_{g} = 1/2 - 0.29 \pm 0.07 
= 0.21 \pm 0.07$. From our results 
one can draw the following conclusions:
\\
1. The total angular momentum content of the quark in the proton 
is $J_{q} = 0.29 \pm 0.07$, {\it i.e.,} 
 about 60\% of the proton spin is contributed by the quarks. Since the lattice result
in ref.~\cite{liu} gives $\Sigma = 0.25 \pm 0.12$, one can deduce
that the quark orbital angular momentum is $0.17 \pm 0.08$.
Thus, one concludes that about 25\% of the
proton spin comes from the quark spin and about 35\% comes from the
quark orbital angular momentum. 
\\
2. Gluon angular momentum contribution is predicted to be 
$J_{g} = 0.21 \pm 0.07$, {\it i.e.}, about $\sim$ 40\% of the 
proton spin is attributable to the glue.
\\
3. 
For the sea quarks, 
almost all the contribution 
comes from the quark spin and the total sea quark contribution 
nearly cancels the contribution due to glue.
As a result,
the proton spin is almost saturated by the CI (valence \& cloud quark contribution) of the quark angular momentum alone.
  
This work is partially supported by DOE Grant No. DE-FG05-84ER40154, DE-FG02-88ER40448 and NSF Grant No. 9722073.

\end{document}